# An interpretive conjecture for physics beyond the standard models: generalized complementarity

Gilles Cohen-Tannoudji[*]

Laboratoire de Recherche sur les Sciences de la Matière (LARSIM)
CEA-Saclay

**Abstract**

Our interpretive conjecture is inspired by the epistemology due to Ferdinand Gonseth (1890-1975) who interpreted complementarity as the relationship between profound and apparent reality horizons. It consists, on the one hand, on enlarging the scope of quantum theory to the most profound reality horizon, namely a triply quantum theory of gravitation that would be able to take into account simultaneously as elementary quanta the Planck's constant, the Planck's space-time area and the Boltzmann constant, and, on the other hand, on interpreting in terms of generalized complementarity three doubly quantum schemata taking into account by pairs, these three elementary quanta and form the apparent reality horizon.

## 1/ Introduction

The recent advances in particle physics with the discovery of the last missing link of its standard model, the Higgs boson, and in cosmology with the establishment of a new standard cosmological model, the so-called ΛCDM model, offer a rare opportunity of assessing the current status of the philosophy of physics and of proposing conjectures about its future in connection with the most challenging issue of elaborating a quantum theory of gravitation.

In a famous article written in 1936 [1] , Albert Einstein stresses how difficult it seems to him to reconcile relativity (special and general) with quantum theory;

> "This leads us to transpose the statistical methods of quantum mechanics to fields that is to systems of infinitely many degrees of freedom. Although the attempts so far made are restricted to linear equations, which, as we know from the results of the general theory of relativity, are insufficient, the complications met up to now by the very ingenious attempts are already terrifying. They certainly will raise sky high if one wishes to obey the requirements of the general theory of relativity, the justification of which in principle nobody doubts."

---

[*] Gilles.Cohen-Tannoudji@cea.fr

While the "terrifying complications" have already been overcome through the quantum field theory used in the physics of non-gravitational fundamental interactions, the difficulties raising "sky high" of reconciling general relativity with quantum theory, that is of elaborating a quantum theory of gravitation are still facing us. However we believe that some progress is currently being made towards their solution, thanks to the very interesting heuristic schemata that we shall discuss below.

The epistemology which underlies our conjecture of generalized complementarity is due to the Swiss mathematician-philosopher *Ferdinand Gonseth* (1890-1975). In a contribution [2] to a special issue, devoted to the concept of complementarity, of the review *Dialectica* he interpreted complementarity as a schema articulating two *reality horizons*: an *apparent* one which is the realm of experiment, intuition and phenomenology, and a *profound* one which is the realm of fundamental theoretical issues. In *Dialectica*, the article of Gonseth written in French is accompanied by the following summary in English:

> The concept of complementarity is here presented in a manner both general (abstract) and elementary, suitable for physics as for other fields. Knowledge progresses through an unveiling of successive horizons of reality, three of which are here considered: The natural horizon (Eigenwelt), the classical horizon and the horizon of quantum theory. The concept of complementarity appears at a fundamental level in the relationship between any two horizons of which one plays the role of an apparent, the other of a deeper horizon. The "compulsory" framework for considerations of this kind is a methodology of open dialectic. Within this framework, the concept of "wave-corpuscle" complementarity appears as a specialization of a more general scheme.

The special complementarity issue of the Dialectica review is particularly interesting for the philosophy of physics because it contains the contributions of five of the father-founders of quantum theory (all Nobel prize laureates), W. Pauli (who coordinated, with F. Gonseth, the director of the review, the edition of the issue), N. Bohr, A. Einstein, W. Heisenberg and L. de Broglie. The author of the present paper has analyzed [3]the article of Gonseth in a website devoted to the analysis of articles useful for the history and philosophy of sciences, and has discussed the relevance of the concept of horizon of reality in a survey of the French contributions to the philosophy of sciences [4].

In terms of horizons of reality, our interpretive conjecture consists

1. On deepening the quantum horizon of reality to a *triply quantum horizon*, the one of quantum gravity, involving, together with the elementary quantum of action $\hbar$, two other elementary quanta: the Boltzmann's constant $k_B$, interpreted, when multiplied by Log2, as an elementary quantum of information and the Planck scale surface area [5] $A_P = \frac{\hbar G}{c^3}$, interpreted as an elementary quantum of space-time area

2. On articulating through what we call *generalized complementarity* this most profound horizon with three *doubly quantum horizons*, that can play with respect to it the role of apparent horizons of reality:

   a. *entanglement thermodynamics* taking into account the quantum of action and the quantum of information,

   b. *holography and thermodynamics of horizons* taking into account the quantum of information and the quantum of space-time area,

   c. *gauge/gravity duality* taking into account the quantum of action and the quantum of space-time area.

## 2/ Entanglement thermodynamics relating action and information quanta

### *The quantum of information*

Quantization of information is advocated by all approaches which give information a foundational role. Usually, the necessity of quantizing information has been tackled through the "exorcism" of the Maxwell demon, this allegory or thought experiment imagined by Maxwell, in which a "demon" would be able to sort the molecules of a gas at thermal equilibrium according to their velocities and to create, without any expense of work, a difference of temperature, thus violating the second principle of thermodynamics (see [6] for a historical survey). According to Wheeler [7], the most convincing exorcism lies in the Landauer [8] principle: "The thermodynamically costly act, which prevents the demon from breaking the second law, is not (as often supposed) the measurement by which the demon acquires information about the molecule being sorted, but rather the resetting operation by which this information is destroyed in preparation for making the next measurement." The smallest amount of information which has to be destroyed in this operation, equal to *k*Ln2, is the quantum of information; it can be called the quantum of *relevant information*, according to the terminology of [9].

### *The Unruh effect and entanglement thermodynamics in quantum field theory*

If action and information are quantized, we expect the ratio of the Planck's and Boltzmann's constants to express, in the profound horizon, their equivalence. In a relativistic framework, the *acceleration radiation* effect discovered by Unruh [10] precisely involves this ratio. This effect has been carefully analyzed by Paul Teller [11]. Ordinary quantum field theory is formulated in a flat (Minkowski) space-time and it is assumed that all observers move inertially. Teller calls "Minkowski quanta" the particles that are registered by a detector moving inertially. Now it turns out that a detector that registers no particles when moving inertially in the vacuum state, which Teller calls the "Minkowski vacuum", registers particles, that he calls "Rindler quanta", when it is in a uniformly accelerated motion. Actually, the Rindler quanta are the ones of a thermal bath with a temperature $T$ related to the acceleration $a$ by means of the "Unruh constant": $U = \frac{\hbar}{k_B c}; T = U \frac{a}{2\pi}$. It is important to note that the phenomenon of black hole evaporation conjectured by Hawking [12] is nothing but an acceleration radiation effect with acceleration equal to the surface gravity at the black hole horizon, and temperature related to it by means of the Unruh constant: $T = \frac{\hbar}{kc} \frac{c^4}{8\pi GM}$.

In [13] the authors stress the difference between ordinary statistical thermodynamics and what they call *entanglement thermodynamics*. Quantum entanglement induces entropy even in a pure state of a pair of entangled degrees of freedom living on both sides of a horizon when one integrates out one of the two degrees of freedom. The thermal bath just above mentioned of Rindler quanta seen by an accelerated observer and not seen by an inertial observer relate *to entanglement thermodynamics and not to statistical thermodynamics*: whereas in statistical thermodynamics, entropy measures our ignorance, in entanglement thermodynamics, entropy measures quantum entanglement, namely a "genuinely quantum phenomenon which does not exist in classical physics". Let us note to conclude this section that S. Kolekar and T. Padmanabhan have established [14] the *Indistinguishability of thermal and quantum fluctuations*, which confirms that the Boltzmann constant has to be considered as a genuine elementary quantum equivalent to the quantum of action.

## 3/ Holography and thermodynamics of horizons

### The Bekenstein constant and the black hole entropy

Thermodynamics needs not only the possibility of defining a temperature but also of defining an entropy. The Unruh constant allows defining a temperature on a Rindler horizon and it is the work of Bekenstein and Hawking on the thermodynamic interpretation of the dynamics of black holes that allows defining entropy on horizons. We sketch here the reasoning that led Bekenstein [15] to associate to the horizon of a black hole an entropy proportional to the area of the event horizon expressed in Planck's units. Let us assume that we drop into the black hole of mass *M* the smallest amount of energy capable to carry the minimal amount of information, namely the quantum $k_B \mathrm{Ln} 2$, in the form of a photon, the wave length of which is the Schwarzschild radius of the black hole $R_S = 2MG/c^2$. The increase of the energy content of the black hole is $\Delta Mc^2 = \Delta E = \hbar c / R_S$, so that $\Delta M = \hbar /(cR_S)$, which implies that the radius is increased by $\Delta R_S = 2\Delta MG/c^2 = 4\pi \hbar G/(c^3 R_S)$ and the area is increased by $\Delta A_H = 8\pi \Delta R_S R_S = 4\hbar G/c^3$, which is a universal constant equal to four times the Planck's area. From this differential evaluation we can extrapolate to the total entropy of the black hole by assuming that a small black hole has zero entropy, so that there is no integration constant. We thus expect $S_{BH} = \eta k_B \mathrm{Ln} 2 \frac{1}{4A_P} A_{BH}$, where the unknown constant $\eta$ is of order 1. Actually, from the Hawking [16] temperature, it is easy to fix the value of this unknown constant, and one finally gets $S_{BH} = B A_{BH}$, where the constant $B = \frac{k_B}{4A_P}$ (*B* for 'Bekenstein') translates the equivalence principle between information and surface area.

### *The Bekenstein bound and the holographic principle*

Since forming a black hole is the most efficient way to compress matter inside a given volume, the Bekenstein entropy appears as a bound on the information content of any sphere in space-time. This leads to the *holographic principle* [17]: "How many degrees of freedom are there in nature, at the most fundamental level? The holographic principle answers this question in terms of the area of surfaces in space-time: (…) A region with boundary of area *A* is fully described by no more than *A*/4 degrees of freedom, or about 1 bit of information per Planck area."

### *Holography and gravity-thermodynamics connection*

The idea that gravity can be described as an emergent phenomenon has a very long history starting from the work of Sakharov [18]. The gravity-thermodynamics connection was discovered by Ted Jacobson [19] who used the proportionality of entropy and the horizon area and classical thermodynamic identities to derive the Einstein's equation as an equation of state. The implications of this connection have then been thoroughly and comprehensively investigated by Padmanabhan [20] and recently E. Verlinde also advocated it by interpreting gravity as an entropic force [21]

In [22] Padmanabhan had shown that general covariance and the principle of equivalence force us to include in the semi-classical action functional for quantum gravity a term involving the second derivative of the metric. This can be done by adding to the bulk term in the Lagrangian a surface term, depending on the second derivative of the metric that can be ignored in a classical theory but not if one wants to quantize the theory of gravity: to have the correct symmetry, the Lagrangian must be of the form $L \propto (\partial g)^2 + \partial^2 g \equiv L_{bulk} + L_{sur}$. If, to describe the dynamics, one uses only $L_{bulk}$, which means that one assumes that the degrees of freedom are the components of the metric and that they reside in the space-time volume, one obtains a description that is highly gauge-redundant. Actually, since one can choose around any event, a local inertial frame, so that $L_{bulk} \propto (\partial g)^2$ vanishes, whereas one cannot make the surface term to vanish by any choices of coordinates, it seems that "the true degrees of freedom of gravity for a volume $\mathcal{V}$, which cannot be eliminated by a gauge choice, reside in its boundary $\partial\mathcal{V}$ and contribute only to $L_{sur}$ around any event".

Furthermore it turns out that the bulk and the surface parts of the Einstein Hilbert Lagrangian are related by a relation that shows that both have the same information content

$$\sqrt{-g}L_{sur} = -\partial_a \left( g_{ik} \frac{\partial \sqrt{-g}L_{bulk}}{\partial(\partial_a g_{ik})} \right)$$

This relation shows that the transition from $L_{bulk}$ to the Einstein-Hilbert lagrangian $L_{EH} = L_{bulk} + L_{sur}$ is very similar to the transition, in Lagrangian mechanics, from the co-ordinate representation characterized by $L_q(\dot{q}, q)$ to the momentum representation characterized by $L_p = L_q - \frac{d}{dt}\left(q \frac{\partial L_q}{\partial \dot{q}}\right)$ which give the same equations of motion.

### *Einstein equation and horizon thermodynamics*

The role of the surface term in determining the true degrees of freedom is supported by the evaluation of the area law of the horizon entropy. Consider a spherically symmetric metric with a horizon: $g_{00} = 1/g_{rr} = -f(r)$ where the function $f(r)$ vanishes at the horizon $r = a$. Near the horizon, the metric reduces to the Rindler metric with a surface gravity (acceleration) equal to $\kappa = (c^2/2)f'(a); (f'(a) \neq 0)$ The discussion in the preceding section allows identifying a temperature associated with the horizon $T = \dfrac{\hbar c f'(a)}{4\pi k_B} = \dfrac{1}{2\pi}\dfrac{\hbar}{k_B c}\kappa$, where we recognize what we called the Unruh constant. We can now write the Einstein's equation for this metric in the form $(1-f) - rf'(r) = -(8\pi G/c^4)Pr^2$ where $P = T_r^r$ is the radial pressure. On the horizon $r = a$, the equation reduces to $\dfrac{c^4}{G}\left[\dfrac{1}{2}f'(a)a - \dfrac{1}{2}\right] = 4\pi P a^2$. We now consider two solutions of the Einstein's equation differing infinitesimally at horizons of radii $a$ and $a + da$, and multiplying by $da$, we get $\dfrac{c^4}{2G}f'(a)ada - \dfrac{c^4}{2}da = P(4\pi a^2 da)$, and multiplying and dividing the first term by $k_B$ and by $\hbar$ (in order to make appear the Unruh and the Bekenstein constants) one obtains the following interpretation of the Einstein's equation as an entanglement thermodynamic identity for the infinitesimal displacement of the horizon of an amount $da$

$$\underbrace{\dfrac{\hbar}{k_B c}\dfrac{c^2 f'(a)}{4\pi}}_{T}\underbrace{\dfrac{k_B c^3}{G\hbar}d\left(\dfrac{1}{4}4\pi a^2\right)}_{dS} - \underbrace{\dfrac{1}{2}\dfrac{c^4 da}{G}}_{-dE} = \underbrace{Pd\left(\dfrac{4\pi}{3}a^3\right)}_{PdV}$$

### *Towards a solution to the cosmological constant problem?*

Following his investigations along this line of thinking, T. Padmanabhan reinterprets the standard equations of Cosmo-dynamics in terms of a *holographic principle of equipartition* [23] underlying the *emergent perspective of gravity* [24] and eventually leading to a possible solution to the cosmological constant problem [25].

## 4/ Gauge/gravity duality, quantum field theory and quantum gravity

### *A quantum extension of the equivalence principle*

The last doubly quantum reality horizon that we now discuss relates the quantum of action to the quantum of space-time area, the ratio of which is nothing but the inverse of the Newton constant *G*. Since all non-gravitational interactions are described by gauge theories, gauge/gravity duality is reminiscent of the classical equivalence principle which states that in absence of non-gravitational forces a test body is not submitted to gravitation, and somehow extends this principle to quantum gravity.

### *Hadronic strings: the prehistory of gauge/gravity duality*

In the seventies the hadronic string heuristic model picturing hadrons as open strings the end points of which are confined quarks or antiquarks provided a hint toward a sub-hadronic confining theory. When QCD was discovered, this picture became more ambitious following the work of G. 't Hooft [26] who conjectured that at the limit when the number of colors $N_c$ goes to infinity with $g^2 N_c$ fixed ($g^2$ being the squared of the QCD coupling constant), QCD confines quarks and antiquarks as the end points of color singlet hadronic strings. Then G. Veneziano, considering the limit where both the number of colors and the number of flavors go to infinity showed that when unitarizing the hadronic string picture, emerges beside the open string picture a *closed string sector*, involving purely gluonic hadronic strings. But at that time the string picture was being abandoned to the profit of QCD to describe hadrons at the fundamental level and to the profit of a theory of fundamental strings at the Planck scale to describe quantum gravity, with open strings for matter and closed strings for gravitation.

### *Gauge/gravity duality in string theory*

After huge developments in research about string theory applied to quantum gravity, the interest to string theory came back to the research of the string theories equivalent to QCD, and that is how emerged the concept of gauge/gravity duality through the so-called ADS/CFT (Anti de Sitter/Conformal Field Theory) concept [27]. G.T. Horowitz and J. Polchinsky begin their introductory paper [28] on gauge/gravity duality by the following assertion: "Hidden within every non-Abelian gauge theory, even within the weak and strong nuclear interactions, is a theory of quantum gravity".

### *Gauge/gravity duality in quantum field theory*

These authors explain that the path to this duality has been tortuous but they suggest that it can also be uncovered in the framework of quantum field theory. In a pedagogical review article [29], S.L. Adler explains how the Einstein Hilbert gravitational action is obtained as a symmetry-breaking effect in quantum field theory. He shows that for a non-abelian gauge theory with massless matter fields the scale invariance can be dynamically broken due to the singular behavior of the theory in the infrared, and he shows that the coefficients of the Einstein Hilbert lagrangian (at least the cosmological constant) can in principle be computed: the Nambu-Goldstone scalars of the symmetry breaking (the longitudinal component of the intermediate vector bosons in the case of the Brout Engler Higgs mechanism in the electroweak theory, and the pions in the case of chiral symmetry breaking in QCD) have a "dilaton" partner (the Higgs boson in electroweak theory and the sigma boson in QCD), namely a massless scalar with quantum numbers of the vacuum, that acquires mass in the symmetry-breaking mechanism, that gives mass to the particles with which it couples, with a coupling proportional their mass. This sounds like a scalar mediated gravitational interaction inducing a cut-off in the infrared singular non-abelian gauge theory. Could it be that this gravitational interaction would be nothing but quantum gravity induced through gauge/gravity duality at the scales of the infrared cut-offs?

## 5/ Conclusion

We have concluded the section devoted to gauge/gravity duality as well as the preceding one devoted to holography by questions that we leave open because their answers relies on a lot of work by researchers with much more expertise than ours. In any case, we want to stress in conclusion that our conjecture of generalized complementarity seems to have a fairly good heuristic power: in each doubly quantum reality horizon it is at work by relating the two quanta which are equivalent in the doubly quantum reality horizon (considered as the profound horizon) and contradictory in the singly quantum reality horizons (considered as forming the apparent horizon); in the same way, generalized complementarity leads one to consider the two schemes discussed in sections 3 and 4 (holography and gauge/duality) as equivalent in the most profound reality horizon, the one of the triply quantum gravity, and contradictory in the pair of doubly quantum reality horizons forming the apparent horizon.

**Acknowledgements** It is a pleasure to acknowledge Michel Spiro, my co-author of the book we wrote to explain to a wide audience the importance of the discovery of the BEH boson [30], and very fruitful discussions with Alexei Grinbaum, Carlo Rovelli, Vincent Bontems and Etienne Klein.